\documentclass[a4paper,11pt]{article}
\pdfoutput=1 

\usepackage{jinstpub} 

\title{\boldmath Performance studies of MRPC prototypes for CBM}

\author[a]{I. Deppner,\note{Corresponding author.}}
\author[a]{N. Herrmann,}
\author[b]{J. Fr\"uhauf,}
\author[b]{M. Ki\v{s},}
\author[c]{P. Lyu,}
\author[b]{P.-A. Loizeau,}
\author[c]{L. Shi,}
\author[a]{C. Simon,}
\author[c]{Y. Wang,}
\author[c]{B. Xie}

\affiliation[a]{Physikalisches Institut der Heidelberg University, Heidelberg, Germany}
\affiliation[b]{GSI Helmholtz Center for Heavy Ion Research GmbH (GSI), Darmstadt, Germany}
\affiliation[c]{Department of Engineering Physics, Tsinghua University (THU), Beijing, China}

\emailAdd{deppner@physi.uni-heidelberg.de}

\abstract{Multi-gap Resistive Plate Chambers (MRPCs) \cite{CerronZeballos:1995iy} with multi-strip readout are considered to be the optimal detector candidate for the Time-of-Flight (ToF) wall in the Compressed Baryonic Matter (CBM) experiment. In the R\&D phase MRPCs with different granularities, low-resistive materials and high voltage stack configurations were developed and tested. Here, we focus on two prototypes called HD-P2 and THU-strip, both with strips of 27 cm$^2$ length and low-resistive glass electrodes. The HD-P2 prototype has a single-stack configuration with 8 gaps while the THU-strip prototype is constructed in a double-stack configuration with 2 $\times$ 4 gaps. The performance results of these counters in terms of efficiency and time resolution carried out in a test beam time with heavy-ion beam at GSI in 2014 are presented in this proceeding.}

\keywords{Gaseous detectors, Resistive-plate chambers, Timing detectors and high counting rate}

\proceeding{XIII$^{\text{th}}$ Workshop on Resistive Plate Chambers and Related Detectors\\
  22 - 26 February, 2016\\
  Gent, Belgium}

\begin{document}
\maketitle
\flushbottom

\section{Introduction}
\label{sec:intro}
CBM \cite{Senger:2012wr} is a future fixed target heavy ion experiment located at the Facility for Antiproton and Ion Research (FAIR) in Darmstadt, Germany. CBM aims to study the phase diagram of strongly interacting matter at the highest baryon densities with unprecedented accuracy. This goal can be achieved by measuring particle yields and flow of very rare probes like multi-strange objects but also bulk particles like pions, kaons and protons. Therefore, an excellent particle identification (PID) at a targeted interaction rate of 10 MHz is required. The PID of charged hadrons at CBM is realized by a Time-Of-Flight (ToF)-wall \cite{Deppner2012} positioned at about 10 m from the interaction point which is composed of Multi-gap RPCs with a total active area of about 120 m$^2$. The requirements of the full system is a time resolution better than 80 ps at an efficiency higher than 95 \%. Since the charged particle flux drops exponentially from small angle (2.5$^\circ$) to large angle (25$^\circ$), ranging from 25 kHz/cm$^2$ to 1 kHz/cm$^2$ high rate capable MRPCs with low-resistive electrode material \cite{Naumann:2011, Wang:2010bg, Wang:2014tsa, Wang:2016bsx, Petrovici:2012dc, Petris:2012zz, Petris:2016rvv} but also with thin float glass \cite{Deppner:2012td, Simon:2014pta} and with different granularities were developed. An overview of the conceptual design of the CBM ToF-wall is given in \cite{Deppner:2014sua}.\\

\section{Counter and experimental setup configuration}

\begin{figure}[htbp]
\centering 
\includegraphics[width=.8\textwidth,origin=c]{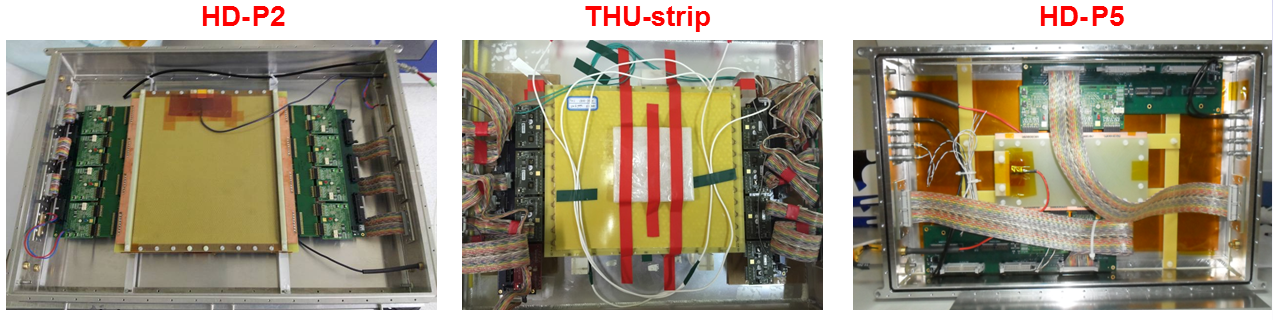}
\caption{\label{fig:1} Photograph of the counters under test (HD-P2 and THU-strip) and the
  reference counter (HD-P5) embedded in their gas-tight box. The preamplifiers
are connected directly to the readout electrode.}
\end{figure}

For the intermediate rate region (between 1.5 and 10 kHz/cm$^2$) in the current conceptual design we consider differential MRPCs with low-resistive glass of 0.7 mm thickness as floating electrodes and multi-strip readout electrodes with a granularity of 27 cm$^2$, i.e. a strip length of 27 cm and a strip pitch of 1 cm (the strip width is 7 mm). MRPCs with such a readout electrode geometry have typically an impedance of about 100 $\Omega$ in a single-stack configuration with 8 gaps and about 50 $\Omega$ impedance in a double-stack configuration with 2 $\times$ 4 gaps.\\ 
In order to evaluate the importance of impedance matching to the front-end electronics which has in our case 100 $\Omega$ a single-stack prototype called HD-P2 and a double-stack prototype called THU-strip with the properties listed in table \ref{tab:1} were built. As reference, a smaller counter called HD-P5 was used. Figure \ref{fig:1} shows a photograph of all three counters.

\begin{table}[htbp]
\centering
\caption{\label{tab:1} Counter types and their properties.}
\smallskip
\begin{tabular}{|l|c|c|c|}
\hline
Counter types & HD-P2 & THU-strip & HD-P5\\
\hline
Active area & 32 $\times$ 27 cm$^2$& 24 $\times$ 27 cm$^2$  & 15 $\times$ 4 cm$^2$\\
\# strips & 32 & 24 & 16\\
Strip width / gap & 7 mm / 3 mm & 7 mm / 3 mm  & 7.6 mm / 1.8 mm \\
Glass thickness & 0.7 mm & 0.7 mm  & 1 mm\\
Glass stack & single & double & single\\
\# gaps & 8 & 2 $\times$ 4& 6\\
Gap width & 220 $\mu$m & 250 $\mu$m & 220 $\mu$m\\
\hline
\end{tabular}
\end{table}

\begin{figure}[htbp]
\centering 
\includegraphics[width=.8\textwidth,origin=c]{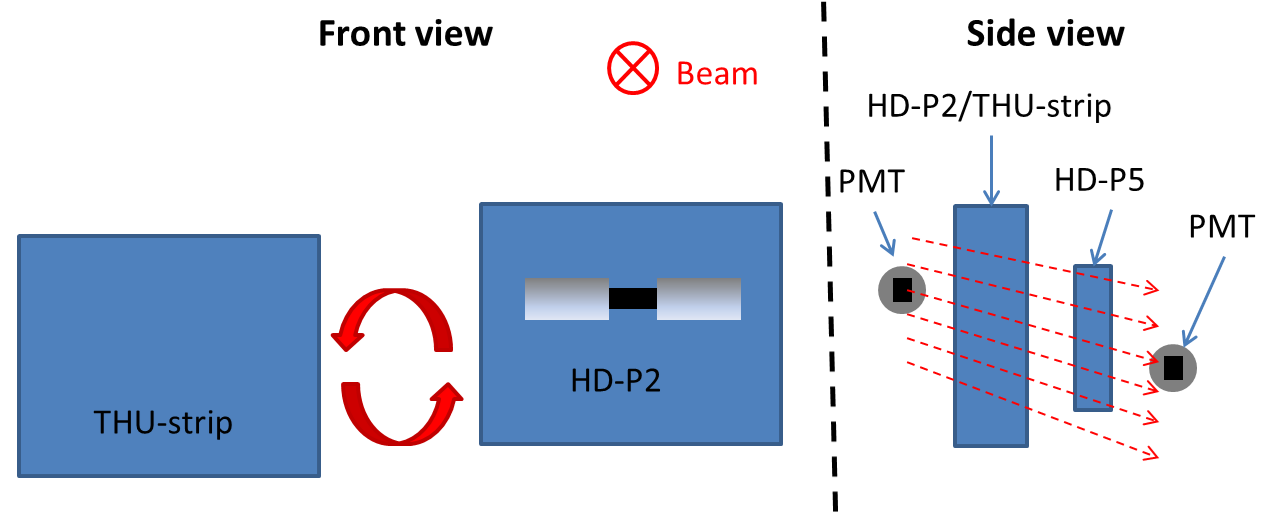}
\caption{\label{fig:2} Low-rate part of the in-beam test setup. The counters were placed
below the beam. The distance to the target is about 5 m.}
\end{figure}

These counters were tested the first time in October 2014 at GSI with reaction products stemming from a samarium beam with a kinetic energy of 1.2~GeV impinging on a 5 mm thick lead target creating a mean particle flux of a few hundred Hz/cm$^2$. The particle flux was measured by two plastic scintillators arranged in front and behind a stack of two MRPCs containing the detector under test (Dut) and the reference counter. The HD-P2 prototype was replaced after some testing time by the THU-strip. This setup (see fig. \ref{fig:2}) was part of a bigger setup including other MRPC counters which were tested in parallel \cite{Petris2016}. Also part of the setup was a diamond detector placed in the beam a few cm in front of the target. 
The setup was positioned about 5 m from the target ensuring a quite uniform illumination on the counter surfaces. The electronic chain components consisted of a preamplifier discriminator stage called PADI6 \cite{Ciobanu2014}, an FPGA based TDC \cite{Ugur2013} and an FPGA based readout system called TRB3 \cite{TRB3}. The gas mixture was composed of 85\% $C_2H_2F_4$, 10\% $SF_6$ and 5\% iso-$C_4H_{10}$.     

\section{Results}

The data are processed in three major steps. In step 1, the data are unpacked and TDC values are calibrated with respect to the differential non-linearities. In step 2, the time information of both strip ends is combined to form a hit with a given position and a mean time. The hit position offsets (which are in fact time offsets) are obtained by shifting the mean of the hit distributions of all strips to zero (position alignment). The same procedure is done for the time offset with respect to the reference counter. Subsequently, the Time-over-Threshold (ToT) value is calibrated by shifting the mean of the ToT distribution to a constant value. In addition, a gain is introduced in order to have a common spread in the ToT distributions. After this procedure the slewing correction with the normalized ToT is applied. The last action in step 2 is the clusterization where hits with similar positions and arrival times are grouped to a cluster. The time and position of every cluster is calculated by taking the mean of the times and positions of the contributing hits and weighting them with their ToT. The same calibration is done for the reference counter. The clusters from both detectors of a single event are matched to proper pairs (again in position and time). All procedures of step 2 are done iteratively until no changes in position and time occur. All correction parameters are stored in a ROOT file. Step 3 is the data analysis. First, corrections on the velocity spread of particles are applied by taking the diamond counter into account. Now cuts on the reference counter can be applied to the cluster position (in order to prevent edge effects), to the cluster multiplicity (multi-hit study), to the cluster matching in-between the counters (usually the best matching pair is used), to the velocity of the particles and so on. 

\subsection{Efficiency}

\begin{figure}[htbp]
	\centering 
	\includegraphics[width=.8\textwidth,origin=c]{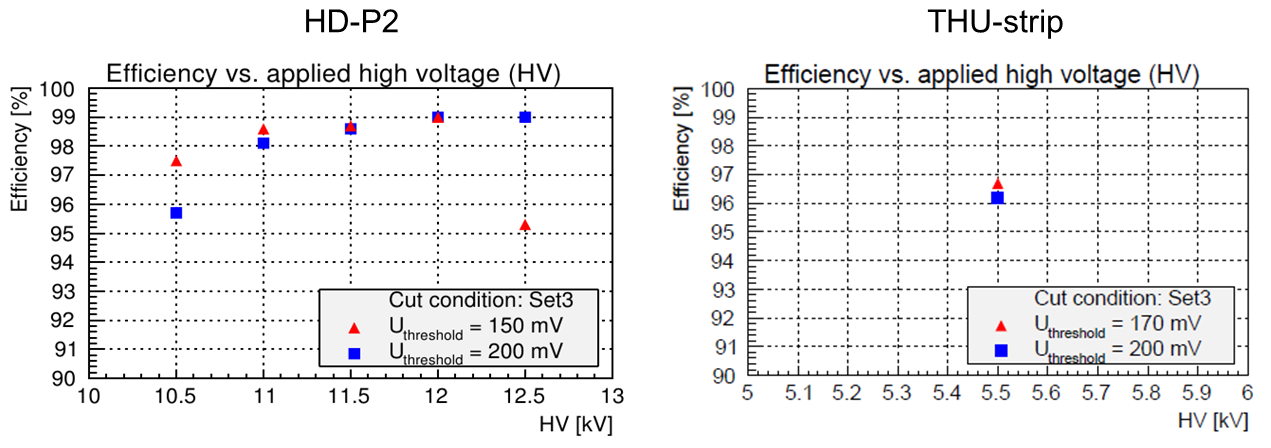}
	\caption{\label{fig:4} Efficiency vs. applied high voltage.}
\end{figure}

The efficiency is calculated by taking a track formed by the diamond and the reference counter and checking if the Dut has a cluster in the vicinity of the penetration point. This can be done since the reference counter is smaller than the Dut. In order to be consistent with the time resolution results, cuts were applied on the position of the clusters on the reference counter. The efficiency vs. applied high voltage for the HP-P2 (left) and THU-strip (right) for two different preamplifier thresholds is shown in fig. \ref{fig:4}. For the HD-P2 counter a successful voltage scan was performed. At 10.5 kV (corresponding to an electric field of about 120 kV/cm) the counter is still not in the plateau. The plateau is reached at about 11 kV (125 kV/cm) with an efficiency higher than 98 \%. The red data point at 12.5 kV is not understood yet since the error is smaller than 0.1 \% and within the symbols. For the double-stack THU-strip counter the efficiency scan was not successful since the data files were corrupt. Only the value at 5.5 kV (corresponding to an electric field of 110 kV/cm) could be analyzed. The efficiency is about 96 \%. Here we assume that the plateau is reached with the argument that the difference of the efficiency for two different thresholds is only about 0.4 \% and in the same order as for the single-stack at 11 kV. The reason for reaching the efficiency plateau at lower fields is a higher mean weighting field in the double-stack counter. However, with only one data point it is not possible to claim that the maximal efficiency is already reached. 

\subsection{Time resolution}

\begin{figure}[htbp]
	\centering 
	\includegraphics[width=.8\textwidth,origin=c]{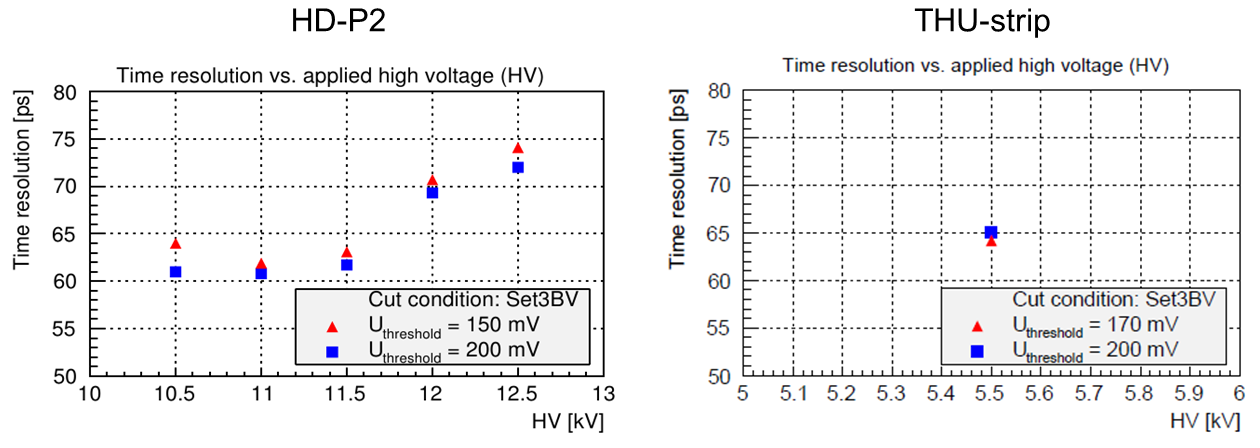}
	\caption{\label{fig:5} System time resolution vs. applied high voltage.}
\end{figure}

The system time resolution versus the applied high voltage for HD-P2 (left) and THU-strip (right) at two different preamplifier threshold settings are shown in fig. \ref{fig:5}. System time resolution means in this context that the time resolution of the reference counter and also the electronics contribution is still included. For the HD-P2 counter we observe a minimum of about 61 ps at about 11 kV followed by a rise probably caused by the onset of streamers. Assuming both counters (Dut and reference) have the same contribution to the system time resolution the single counter resolution is about 43 ps for the best value. Apparently, there is a trend that the system time resolution at the lower threshold is slightly higher. However, the discrepancy is still within errors which are for all data points about 1 ps. In addition, the situation for the THU-strip counter is opposite. The system time resolution for the THU-strip is slightly worse (65 ps) which could be explained by a bigger gap size. However, this speculation could not be tested due to the lack of systematic data. For the data shown in fig. \ref{fig:5} an additional cut is applied with respect to the efficiency plot. The cut excludes the slowest 3 \% of particles (velocity cut), i.e. from a total particle arrival time spread of 10 ns the particles with more than 7 ns delay are excluded. However, taking all particles into account the system time resolution grows only by about 3 ps. 

\begin{figure}[htbp]
	\centering 
	\includegraphics[width=.8\textwidth,origin=c]{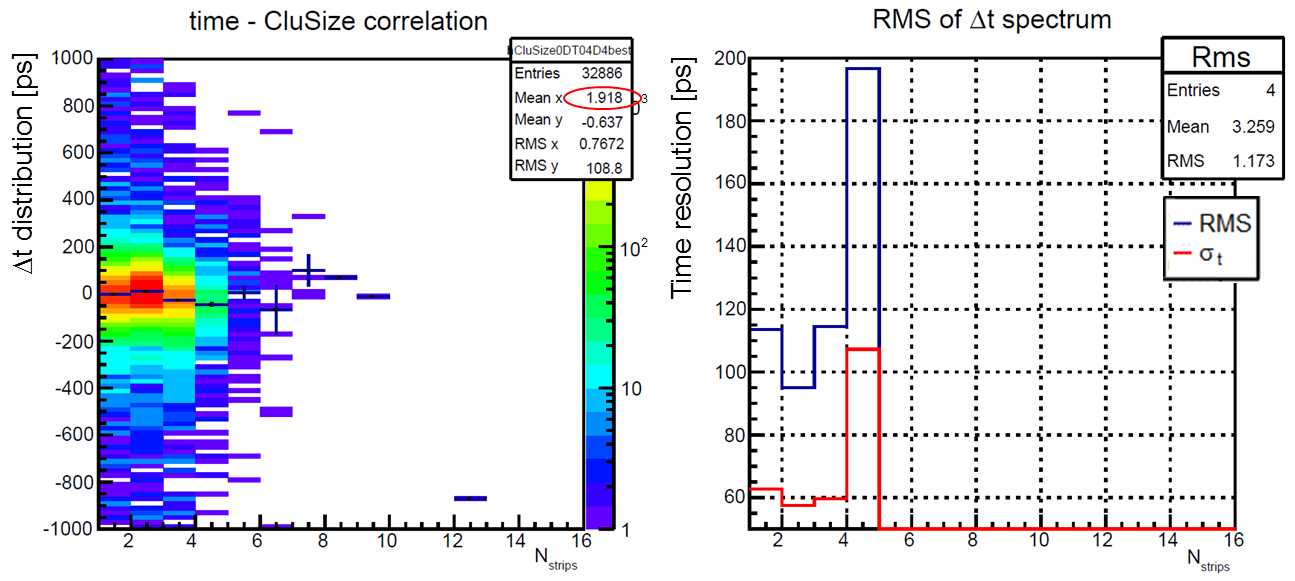}
	\caption{\label{fig:6} Left: time distribution between Dut and reference counter as function of the cluster size of the Dut. Right: RMS value (blue) and the sigma of a fitted Gaussian (red) for each bin.}
\end{figure}

\begin{figure}[htbp]
	\centering 
	\includegraphics[width=.8\textwidth,origin=c]{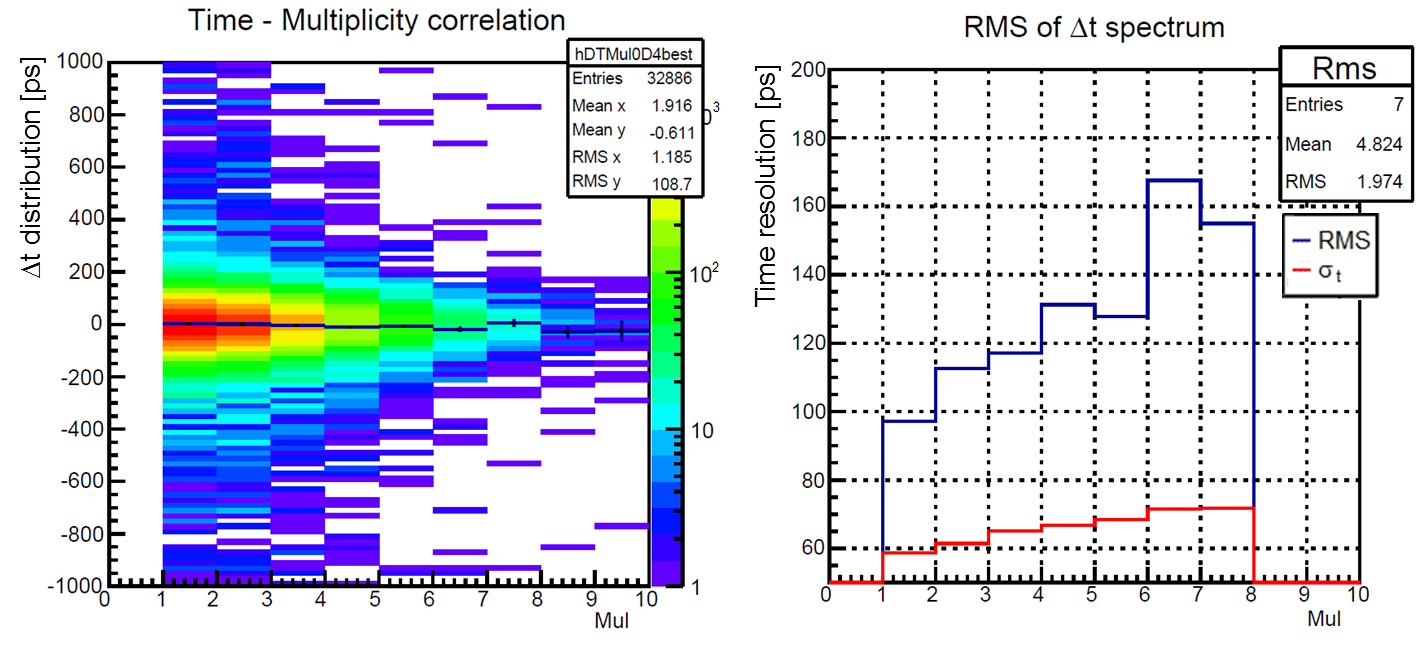}
	\caption{\label{fig:7} Left: time distribution between Dut and reference counter as function of the cluster multiplicity. Right: RMS value (blue) and the sigma of a fitted Gaussian (red) for each bin.}
\end{figure}

Figure \ref{fig:6} shows on the left side the time distribution between HD-P2 and reference counter as function of the cluster size of the Dut. The plot on the right side shows the RMS value (blue) and the sigma of a fitted Gaussian (red) denoting the system time resolution for each bin. 
The mean cluster size is the mean x value (red ellipse) of the 2d plot which is in this case (at a high voltage of 11 kV and a threshold of 200 mV) 1.92. The system time resolution has a minimum at cluster size 2. At cluster size 3 the resolution is still better than at 1 but suffers from lower statistics. This trend can be explained by the fact that we use the cluster mean time to get the resolution and therefore every firing strip contributes to the measured time. On the other hand the more strips fire the lower is the signal/charge on the individual strips diminishing the precision of the measurement. These two effects counterbalance and give a minimum at some intermediate cluster size. This trend was also observed in \cite*{Petris2016}.\\
Another interesting dependence is the time distribution between HD-P2 and reference counter and the system time resolution as function of the cluster multiplicity (see fig \ref{fig:7}). Here, cluster multiplicities up to 9 were observed. The left plot showing the RMS values (blue) and the sigma of a fitted Gaussian (red) indicate a constant rise towards the higher multiplicities. However, at multiplicity 7 corresponding to a occupancy of 50 \% a system time resolution of 72 ps ($ \mathrel{\hat=} $ single counter resolution below 50 ps) is observed.

\section{Conclusions}

The performance in terms of efficiency and time resolution is studied for a single-stack counter (HD-P2) and a double-stack counter (THU-strip) in a heavy-ion beam at GSI. For the HD-P2 prototype an efficiency higher than 98 \% and a time resolution below 45 ps as best values were reached. The results of THU-strip are slightly worse even though only data of one high voltage setting are available. In a triggered system with TDC dead times larger than 20 ns, no difference between single-stack and double-stack counters were observed. This might change in a free-running mode where the TDC dead time is below 5 ns and therefore reflections generated by mismatched impedances become visible.

\acknowledgments

This work was partially funded by BMBF grant no. 05P12VHFC7, NASR/CAPACITATI-Modul III, contract nr. 179EU, NUCLEU Project PN09370103 and by the FuturePID work package (WP19) of the HadronPhysics3 activity within the EU's Seventh Framework Program (FP7).



\end{document}